\documentclass[aps,prl,preprint,superscriptaddress]{revtex4}

\begin{document}


\title{On Chern-Simons corrections to  magnetohydrodynamics equations}


\author{Patrick Das Gupta}
\email[]{patrick@srb.org.in}
\affiliation{Department of Physics and Astrophysics, University of Delhi, Delhi - 110 007 (India)}



\begin{abstract}
We study the effect of a (3+1)-dimensional Chern-Simons electrodynamics on the equations governing the dynamics of
magnetized plasma and fields. In this model, the Chern-Simons (CS) part consists of a dynamical pseudo-scalar field whose space-time derivatives
couple with the electromagnetic field. We explore the CS corrections to the evolution equation for the magnetic field in a plasma with
non-zero electrical resistivity. We revisit  Cowling's theorem in this context and observe that the CS corrections lead to possibly small but
non-zero source
terms for axisymmetric magnetic field. The scalar product of electric and magnetic fields play the role of source of the pseudo-scalar field, and therefore, pulsars are
likely astrophysical candidates to generate propagating pseudo-scalar waves. Although aligned electric
field gets shorted out by flowing charges in large parts of the magnetosphere, there are vacuum gaps in the vicinity of pulsars where strong $\vec E. \vec B$ is expected
to be present. We derive
a wave solution for the pseudo-scalar field generated by the time-varying $\vec E. \vec B$ associated with a pulsar. 
\end{abstract}

\maketitle

\section{I. Introduction}
Electromagnetic fields have a dominating influence on plasma behaviour, and therefore, any modification of Maxwell's equations is likely to have an interesting observable effect on magnetohydrodynamics.
 Chern-Simons theory, in which the Lagrangian density contains the 4-vector potential $A_\mu$ explicitly, with its non-trivial topological aspects has found many applications in physics.   
It is well known that the (2+1)-dimensional Cherns-Simons (CS) theory plays a crucial role in planar physics, particularly in the domain of fractional quantum Hall effect and 
anyonic superconductivity [1,2]. Further more, Witten (1989)  established a connection 
between the mathematical theory of knot and link invariants and CS gauge theory [3]. With the paper of Carroll, Field and Jackiw (1990) (CFJ90) it became clear that not only one can
  formulate a (3+1)-dimensional CS electromagnetic theory in terms of a fixed 4-vector $p_\mu$ one can also put constraint on it 
 using radio-data of distant radio-galaxies [4]. 
 
 Thereafter, Jackiw and Pi (2003) (JP03) extended this analysis 
 to construct a (3+1)-dimensional CS-modified theory of gravitation, involving again a fixed vector $p_\mu$ [5]. Setting $p_\mu \propto {{\partial (r \cos \theta)}\over{\partial x^\mu}}$ in a theory based
 on JP03, it was shown that CS term amplifies the gravito-magnetic part of gravity in such a way that speed of a test particle moving in a circular orbit around a blackhole reaches saturation at large distances,
implying the possibility of explaining flat rotation curves observed in galaxies without invoking dark matter [6]. The CS term in gravitation also induces a shift in the quantum phase in neutron interferometry
although the CS correction to phase shift is too small to be observed in current experiments [7].

 To account for the observed late-time acceleration of the universe, it was shown that a dynamical scalar-density of weight 1 can be a source of dark-energy that is
 responsible for the acceleration, and that this model entails a natural CS extension of electrodynamics as well as gravitation 
in the (3+1)-dimension, along the lines of CFJ90 and JP03 [8]. In flat space-time, the scalar-density appears as a pseudo-scalar field $\psi $ that
 changes sign under space-inversion and under time-reversal. 

In the present study, we explore the possibility of Chern-Simons modification of MHD equations, and then show explicitly that CS electrodynamics allows  one to estimate the oscillation characterstics
of the pseudo-scalar wave, generated
 by time-dependent $\vec E. \vec B$ in the vacuum gaps of a pulsar magnetosphere, in which the pulsar is taken to be an oblique magnetic dipole rotator. In the following section, we begin with the
 description of a model that incorporates CS modification of Maxwell's electromagnetism.

\section {II. Chern-Simons modified magnetohydrodynamics}

Two important tensors in physics  remain invariant under proper Lorentz 
transformations - the Minkowski metric $\eta_{\mu \nu}$ and the totally antisymmetric Levi-Civita pseudo-tensor
$\epsilon_{\mu \nu \rho \sigma}$. The latter, however, changes sign under space inversion as well as under time reversal.
The object $w_{\mu \nu \rho \sigma} \equiv \psi (x) \epsilon_{\mu \nu \rho \sigma}$ is a completely antisymmetric fourth rank tensor under both proper and improper
Lorentz transformations provided $\psi $ is a pseudo-scalar. 
By supposing that $w_{\mu \nu \rho \sigma}$ or, equivalently,  $\psi $ to be a dynamical field, we had investigated
its role as a dark energy candidate to explain the observed acceleration of the expansion of the universe [8], and the action that was used for $\psi $
 reduces in flat space-time to,
$$S_\psi = \int {\bigg [{{A}\over{\psi}}  \eta^{\mu \nu}  {{\partial \psi}\over{\partial x^\mu}} {{\partial \psi}\over{\partial x^\nu}} + B \psi \bigg ] d^4 x} \eqno(1)$$
where  $A$ and $B$ are parameters having the physical dimensions  M/T and M/(T$\mbox{L}^2$), respectively, while $\psi $ is dimensionless.  

 The pseudo-scalar $\psi $ can also be used in conjunction with an approach described in  CFJ90 to write down a gauge invariant Chern-Simons action,
$$S_{CS}= \mbox {J} \int {\epsilon^{\mu \nu \alpha \beta} F_{\mu \nu} A_\alpha {1\over{\psi}} {{\partial \psi}\over{\partial x^\beta}} \ d^4 x}\eqno(2)$$
where J has the dimension of T/L. If one assumes $p_\beta \equiv  4 J \mbox{c} {{\partial \psi}\over{\partial x^\beta}}$ to be a constant 4-vector, the  action given by eq.(2) reduces to that described
 in CFJ90. From the observed 
correlation between polarization and position angles of radio-galaxies, CFJ90 obtained a constraint $m \leq 1.2 \times 10^{-42}$ GeV (if one takes the
Hubble constant to be $H_0 \approx $ 72 km/sec/Mpc), where $m$ is the mass associated 
 with the Compton wavelength $(p^\beta p_\beta)^{-1/2} $ . As $\psi $ is dynamical in the present context, this constraint has no bearing on our analysis. However, we expect that the magnitude of $\psi $ does 
 not deviate too much from unity, and that the dimensionless CS coupling constant Jc is small.

The full action for the CS modified electromagnetism is given by,
$$ S=  -{{1}\over{16\pi \mbox {c} }} \int {F^{\mu \nu} F_{\mu \nu}  \ d^4 x} -
 {1\over{\mbox {c}}^2} \int {j^\mu A_\mu d^4x} + S_\psi + S_{CS},\eqno(3)$$
where $j^\mu $ is the electric 4-current density associated with standard matter.

By extremizing  $S$ with respect to $A_{\mu}$ and $\psi $, respectively, we obtain the following dynamical equations,
$${{\partial F^{\alpha \beta}}\over{\partial x^\beta}}= -{{4 \pi}\over{\mbox{c}}} j^\alpha + 8\pi \mbox{c J}\  \epsilon^{\mu \nu \alpha \beta} F_{\mu \nu} {1\over{\psi}} {{\partial \psi}\over{\partial x^\beta}} $$
and,
$$\bigg [{1\over{{\mbox{c}}^2}} {{\partial ^2}\over {\partial t^2}} - \nabla^2 \bigg ] \psi - {{1}\over{2 \psi}}  \eta^{\mu \nu}  {{\partial \psi}\over{\partial x^\mu}} {{\partial \psi}\over{\partial x^\nu}} = {B\over{2A}} \psi  + 
{{\mbox{J}}\over{4 A}}\epsilon^{\mu \nu \alpha \beta} F_{\mu \nu} F_{\alpha \beta}$$
In terms of electric and magnetic fields, the above equations take the following forms,
$$\nabla . \vec E= 4 \pi \rho_e + 16 \pi \mbox{c} \ \nabla g. \vec B \eqno(4a)$$
$$\nabla . \vec B = 0 \eqno(4b)$$
$$\nabla \times \vec B = {1\over{\mbox{c}}} {{\partial \vec E}\over{\partial t}} + {{4 \pi}\over{\mbox{c}}} \vec j - 16 \pi \mbox{c} \bigg [ {1\over{\mbox{c}}}{{\partial  g}\over{\partial t}} \vec B + \nabla g \times \vec E \bigg ]\eqno(4c)$$
$$\nabla \times \vec E = -{1\over{\mbox{c}}} {{\partial \vec B}\over{\partial t}} \eqno(4d)$$
$$\bigg [{1\over{{\mbox{c}}^2}} {{\partial ^2}\over {\partial t^2}} - \nabla^2 \bigg ] f = {B\over{4 A}} f - {\mbox{J}\over{A}} {{\vec E . \vec B}\over {f}}\eqno(4e)$$
where,
$$\psi (\vec r,t) \equiv (f(\vec r,t))^2,\eqno(5a)$$
$$g(\vec r,t) \equiv \mbox {J} \log \psi = 2 \mbox {J} \log f ,\eqno(5b)$$
and $\rho_e$ is the electric charge density.

For any magnetized plasma, where the fluid bulk speed $u$  is non-relativistic, the magnitude of ${1\over{c}} {{\partial \vec E}\over{\partial t}}$ is only of the order ${{u^2}\over{c^2}}$ times
spatial derivatives of magnetic field, so that it
can be neglected in comparison to the other terms in order to arrive at,
$$\nabla \times \vec B =  {{4 \pi \sigma}\over{\mbox{c}}} (\vec E + {{\vec u}\over{\mbox{c}}}\times \vec B) - 16 \pi \mbox{c} \bigg [ {1\over{\mbox{c}}}{{\partial  g}\over{\partial t}} \vec B + \nabla g \times \vec E \bigg ],\eqno(6)$$
after making use of Ohm's law, with $\sigma$ being the conductivity. Taking the curl of eq.(6), we obtain,
$${{\partial \vec B}\over{\partial t}} + \nabla \times (\vec B \times \vec u)  =  -\nabla \times (\eta \nabla \times \vec B) - \nabla \times \bigg [16 \pi \eta \bigg ( {1\over{\mbox{c}}}{{\partial  g}\over{\partial t}}
\vec B + \nabla g \times \vec E \bigg)  \bigg],\eqno(7a)$$
where $\eta \equiv {c^2\over{4 \pi \sigma}}$ is the plasma electrical resistivity. 

As $g$ and its derivatives are expected to be small, while the conductivity of cosmic plasma is in general large, we may neglect $O (g^2)$, $O(g u/c)$ and $O (g/\sigma)$ terms, with the result that,
$$ {{\partial \vec B}\over{\partial t}} + \nabla \times (\vec B \times \vec u)  =  \eta  \nabla^2 \vec B - 16 \pi \eta \bigg [ {{4 \pi }\over{{\mbox{c}}}^2}{{\partial  g}\over{\partial t}}
\vec j + \nabla \bigg ({1\over{\mbox{c}}}{{\partial g}\over{\partial t}} \bigg ) \times \vec B   + 4 \pi \rho_e \nabla g   \bigg] \eqno(7b)$$
While going from eq.(7a) to eq.(7b) we have assumed the conductivity $\sigma $ to be a constant and have used eqs.(4a)-(4c).

We note that even after the above approximations, eq.(7b) still implies $ {{\partial (\nabla.\vec B)}\over{\partial t}} =0$ upto $O(g)$,  for any neutral plasma, since $\rho_e =0$ and thereby $\nabla . \vec j=0$.
Therefore,  dynamical evolution of magnetic field according to eq.(7b) does not result in spurious creation of magnetic monopoles, when $\rho_e=0$. 

 It is interesting to revisit Cowling's theorem [9] in the context of Chern-Simons modified electrodynamics. Extending Shu's approach to Cowling's theorem [10] to study the case of an axisymmetric distribution of plasma matter and fields,  
 we express the magnetic field $\vec B (s,z)$ in circular cylindrical coordinate system $(s, \phi, z)$ as,
 $$\vec B = B_s \hat {e}_s + B_\phi \hat {e}_\phi + B_z \hat {e}_z,$$
  where all the components of $\vec B$ are $\phi $-independent.
  
With $\rho_e=0$, eq.(7b) can be expressed as,
$${{D \vec B}\over{D t}} + \vec B (\nabla .\vec u) - (\vec B.\nabla) \vec u =  \eta \nabla ^2  \vec B -  16 \pi \eta \nabla \times \bigg ( {1\over{\mbox{c}}}{{\partial  g}\over{\partial t}}
\vec B\bigg ),\eqno(8)$$
where,
$${D\over{Dt}} \equiv {\partial\over{\partial t}} + \vec u.\nabla$$
is the comoving derivative along the flow.

Using,
$$s {D\over{Dt}} \bigg ({1\over{s}} \bigg) = - {{u_s}\over{s}},$$
$$ (\nabla^2 \vec B)_\phi = {\partial\over{\partial s}} \bigg ({1\over{s}}{\partial\over{\partial s}}(s B_\phi)\bigg ) + {{\partial^2 B_\phi}\over{\partial z^2}}$$
and,
$$\nabla. \vec u = \rho {D\over{Dt}}\bigg ({1\over{\rho}} \bigg ),$$
which follows from mass conservation, $\rho$ being the mass density of the plasma, the evolution equation (eq.(8)) for the $\phi $ component of the magnetic field takes
the following form,
$${D\over{Dt}}\bigg ({{B_\phi}\over{\rho s}}\bigg ) = {1\over{\rho s}} \bigg [ \eta \bigg \lbrace {\partial\over{\partial s}} \bigg ({1\over{s}}{\partial\over{\partial s}}(s B_\phi)\bigg ) + 
{{\partial^2 B_\phi}\over{\partial z^2}} \bigg \rbrace - 16 \pi \eta \ \bigg (\nabla \times \bigg ({1\over{\mbox{c}}}{{\partial  g}\over{\partial t}}
\vec B \bigg )\bigg )_\phi + B_z {{\partial u_\phi}\over{\partial z}} + B_s \bigg (  {{\partial u_\phi}\over{\partial s}} -  {{ u_\phi}\over{ s}}  
\bigg) \bigg ]\eqno(9)$$

 First term on the right hand side of eq.(9) leads to the usual decrease in $B_\phi $ due to resistive diffusion. The last two terms consisting of poloidal components of magnetic field would also eventually decay [10], 
 and therefore, cannot be an eternal source for $B_\phi$. The second term on the RHS of eq.(9) due to CS correction does raise the possibility of a source or a sink (depending on its sign) for the toroidal component,
 but in normal astrophysical situations,
 its magnitude is expected to be too small to affect
 the conclusion of Cowling's theorem on the impossibility of a steady axisymmetric magnetic field, supported by current distributions. However, eq.(7b) could  possibly play a significant
role in magnetogenesis in the early universe in tiny pockets where fluctuating  $\rho_e$ and $\nabla g$ cause birth of seed $\vec B$ field (also see [11]).

But first of all, is it at all possible to generate time-dependent $g$? We take this up in the next section. 

\section{III. Pulsar as a source of propagating pseudo-scalar wave}
The evidence for strongest magnetic fields  in the universe comes from observations of pulsars and magnetars [12, 13].  Pulsars are rapidly spinning magnetized neutron stars with high values
of angular speed $\Omega $ and  with polar magnetic fields
often exceeding $10^{12}$ G. The associated electric fields near
the surface reach large values causing ions and electrons to be pulled out of the surface of the neutron star [14, 15].  A substantial portion of the stripped ionized matter that are anchored to the closed magnetic field
lines, within the
light-cylinder of radius $\approx \mbox {c}/\Omega $,
form a magnetosphere that largely corotates with the pulsar. Rest of the plasma matter  escape along open field lines as pulsar wind.  Since $\vec E. \vec B$ acts as the source of the pseudo-scalar field
 (eq.(4e)), one needs to investigate the possibility of generating $\psi $ waves from pulsars as well as magnetars (with $|\vec B| > 10^{14}$ G).

In this paper, we adopt the oblique rotator model in which the rotating neutron star is associated with a magnetic dipole moment $\vec \mu (t) $ that maintains an angle
$\alpha $ with respect to the axis of rotation, as it rotates about it. We assume a geometry in which the spin axis of the pulsar is along the z-axis so that its angular velocity is $\Omega \hat {k}$. Then,
the magnetic moment is given by,
$$\vec \mu (t) = {1\over{2}} B_p a^3 [\sin \alpha (\cos(\Omega t) \hat {i}+ \sin (\Omega t) \hat {j}) + \cos \alpha \hat {k}],\eqno(10)$$
where $a$ and $B_p$ are the  neutron star radius and the magnitude of the magnetic fields at the poles, respectively.

For the rotating magnetic dipole moment, the electric and magnetic fields as measured by a stationary observer at the position vector $\vec r$
outside the neutron star but within the light-cylinder are given by [16,17],
$$\vec E = - {\Omega \over{ \mbox{c} r^2}} (\hat {k} \times \vec \mu (t)) \times \hat {n} + {{\Omega a^2}\over{\mbox{c} r^4}} [(\hat {k}.\vec \mu (t)) \hat {n} + (\hat {k} .\hat {n}) \vec \mu (t)
+ (\vec \mu (t). \hat {n}) \hat {k} - 5 (\vec \mu (t).\hat {n})(\hat {k}.\hat {n}) \hat {n}]$$
$$\vec B = {1\over{r^3}} [ 3 (\vec \mu (t). \hat {n})\hat {n} - \vec \mu (t)],\eqno(11)$$
where the centre of the pulsar is taken to be the origin of the coordinate system and $\hat {n}$ is the unit vector along $\vec r$. 
In spherical polar coordinates,
$$\vec \mu (t). \hat {n} = {1\over{2}} B_p a^3 K (\alpha, \theta, \phi)\eqno(12)$$
where,
$$K(\alpha, \theta, \phi) \equiv [\sin \alpha \sin \theta \cos (\Omega t - \phi) + \cos \alpha \cos \theta]\eqno(13)$$
Then, from eqs.(11)-(13),
$$\vec E. \vec B = - {{\Omega B^2_p a^6} \over {4\mbox{c} r^5}} \bigg [ (K \cos \alpha - \cos \theta ) \bigg ( 1- {{a^2}\over {r^2}} \bigg ) + {{4 a^2}\over {r^2}} K^2 \cos \theta \bigg ]\eqno(14)$$
In most parts of the magnetosphere, the electric field along the direction of magnetic field gets shorted out by the conducting plasma. However, the vacuum gap where electrons 
and positrons are accelerated to very high energies by the aligned electric fields are the sites with non-zero $\vec E. \vec B $ [18].
Therefore, the rotating polar and the outer gap regions of the magnetoshere, by virtue of eqs.(4e) and (14), are likely to play a significant role for the generation of propagating $\psi $ field. 

It was shown in [8] that if $\psi $ acts as the source of dark energy (DE)  that leads to  the observed late-time acceleration of the universe, then it is necessary that the value of the
parameter $\sqrt{-{B\over{4 A}}}$ 
is about ${{3 H_0}\over{2\mbox{c}}} \sqrt{\Omega_{DE}}\approx 10^{-28}\  \mbox{cm}^{-1}$, $\Omega_{DE}$ being the present dark energy density parameter. 
Thus, for stellar scales which are  less compared to the Hubble radius $\approx 10^{28}$ cm by a factor of
more than fifteen orders, we can neglect the term $\sqrt{-{B\over{4 A}}} f$ in eq.(4e) so that it simplifies to,
$$\bigg [{1\over{{\mbox{c}}^2}} {{\partial ^2}\over {\partial t^2}} - \nabla^2 \bigg ] f =  - {\mbox{J}\over{A}} {{\vec E . \vec B}\over {f}}\eqno(15)$$
As $f$ is expected to be close to unity, it can be set to,
$$f(\vec r,t)= 1 + h(\vec r,t)\eqno(16)$$
where the magnitude of  $h$ is much less than one. Then, upto $O (J h)$, eq.(15) assumes a simpler form,
$$\bigg [{1\over{{\mbox{c}}^2}} {{\partial ^2}\over {\partial t^2}} - \nabla^2 \bigg ] h =  - {\mbox{J}\over{A}} \vec E . \vec B,\eqno(17)$$
which can be integrated to yield the retarded solution,
$$h(\vec r, t)= - {\mbox{J}\over{4 \pi A}} \int {{{(\vec E.\vec B)(\vec r', t- |\vec r - \vec r'|/c) }\over{|\vec r - \vec r'|}}d^3 r'}\eqno(18)$$
The contribution to the volume integral comes only from the vacuum gap regions where $\vec E.\vec B$ is non-zero. There is considerable uncertainty in the literature
about the structure of vacuum gaps, and the subject of 3-dimensional geometry
of polar and outer gap regions is an active area of research (see, for example, [19,20]). However, we can infer the nature of space-time dependent oscillations of $h$ 
at distances larger than the size of the magnetosphere by substituting,
$$|\vec r - \vec r'| \approx r \bigg (1 - {{\hat {n}.\vec r'}\over{r}} \bigg )\eqno(19)$$
in eq.(18) and then integrating over $\vec r'$, assuming that $\vec E. \vec B$ in vacuum gaps is given by eq.(14).
The resulting expression for $h$ has the following form,
$$ h(\vec r, t) \approx - {{\mbox{J}}\over{16\pi A}} {{\Omega B^2_p a^6}\over{ \mbox {c} r}} [K_1(\alpha) + K_2 (\alpha) \cos (\Omega (t - r/c) + \phi_1) + K_3(\alpha) \cos(2\Omega (t-r/c) + \phi_2)]\eqno(20)$$
where,
$$K_1 (\alpha) \equiv a_1 + a_2 \sin^2\alpha ,$$
$$K_2 (\alpha) \equiv a_3 \sin \alpha \cos \alpha ,$$
$$K_3 (\alpha) \equiv a_4 \sin^2 \alpha, \eqno(21)$$
where the constants $\phi_1$, $\phi_2$, $a_1$, $a_2$, $a_3$ and $a_4$ encapsulate the integrals over the unknown gap geometry, and hence not only depend on the exact 3-dimensional structure
of vacuum gaps but also on $\hat {n}$. The first term in the RHS of eq.(20) represents a static $h$ field falling as ${1\over{r}}$ while the remaining two terms constitute dipole and quadrupole radiation, 
respectively. Hence, the time variation of $h$ occurs with frequencies $\Omega $ and $2 \Omega $. An interesting but difficult problem is to study the propagation of electromagnetic wave through a region
where $h$ is given by eq.(20). 

Up to first orders in $h$, eqs.(5b) and (16) lead us to,
$${{\partial g}\over{\partial t}}\approx 2\mbox {J}{{\partial h}\over{\partial t}} \eqno(22)$$
Hence, using eq.(20) in eq.(22) we may estimate the magnitude of ${{\partial g}\over{\partial t}}$ at a distance of hundred times the light-cylinder radius of a Crab-like pulsar,
$$|{{\partial g}\over{\partial t}}|\approx  5 \times 10^{42} {\mbox{cm}}^{-1}{{\mbox {J}}^2\over{A}} \times \mbox{Geometrical factor} \times  \bigg ({{B_p}\over{10^{12} G}} \bigg )^2 
\bigg ({{a}\over{10\  \mbox {km}}} \bigg )^6  \bigg ({{P}\over{33\  ms}} \bigg )^{-2} \bigg ({{r}\over{10^5\  \mbox{km}}} \bigg ) \eqno(23),$$
with $P\equiv {{2 \pi}\over{\Omega}}$ being the spin-period of the pulsar.
The unknown geometrical factor in eq.(23) arises from the integrations over the vacuum gap region (see eq(21)). Although the numerical figure in eq.(23) is impressive, until we have estimates of the geometrical factor 
 as well as the constants J and $A$,
it is difficult to ascertain the importance of Chern-Simons correction to the evolution equation of the magnetic field (eq.(9)). 

\section{IV. Discussions}
In our model, the Chern-Simons (CS) action originates  due to a coupling between the dynamical fourth rank tensor $\psi (x) \epsilon_{\mu \nu \alpha \beta}$, the electromagnetic field and the 4-vector potential. 
 We derived  the  MHD equations with CS corrections and showed that the latter gives rise
to a source term in the evolution equation for the toroidal component of the magnetic field, assuming an axisymmetric distribution of both a flowing plasma as well as magnetic field. The CS corrections could possibly  lead to
spawning of  seed magnetic
fields in small regions in the early universe. From the dynamical equations, we found that
scalar product of electric and magnetic fields acts as a source of  the pseudo-scalar field $\psi$. Pulsars and magnetars are likely to be strong sources as they 
are associated with large values of  magnetic and electric fields. Although aligned electric field is completely screened in most parts of the magnetosphere of a pulsar, there are
charge-deficit gaps where $\vec E. \vec B$ is expected to be large. We obtained explicit wave 
solutions for $\psi$ generated by time-dependent $\vec E. \vec B$ in the vacuum gaps of a pulsar magnetosphere. 

However, it is to be noted that the vacuum gaps are fairly close to the neutron star where the curvature of space-time geometry cannot be neglected. Strictly speaking, one should solve
the curved space-time equivalent of eq.(4e) (see, for example, [8])
after having determined the background metric in a self-consistent manner using CS gravity around a rotating neutron star.  Although, in recent studies, weak field corrections to the metric tensor arising because of the CS gravity 
term have been computed assuming the pseudo-scalar field to have only time dependence [21,22], it appears that CS gravity does not admit stationary and axisymmetric solution apart from Schwarzschild, flat space-time and two
other 
unphysical solutions no matter how the pseudo-scalar field is chosen [23]. In the absence of a Kerr-like solution in CS gravity, it is difficult to provide a self-consistent calculation of propagating $\psi $ field generated from strong gravity
regions of a pulsar. 

\section{Acknowledgement}
I thank the anonymous referee of this paper for his useful comments and suggestions. 

\section{References}
[1] Ezawa, Z. F. and Iwazaki, A., Phys. Rev. {\bf B 43}, 2637 (1991)

[2] Dunne, G. V., Topological aspects of low dimensional systems, Vol. 69, 1999 Les Houches Lectures (Springer Berlin/ Heidelberg)

[3] Witten, E., Comm. Math. Phys.{\bf 121} 351 (1989)

[4] Carroll, S. M., Field, G. B., and Jackiw, R., Phys. Rev. {\bf D 41}, 1231 (1990)

[5] Jackiw, R. and Pi, S. -Y., Phys. Rev. {\bf D 68}, 104012 (2003)

[6] Konno, K., Matsuyama, T., Asano, Y. and Tanda, S., Phys. Rev. {\bf D 78}, 024037 (2008)

[7] Nandi, K. K., Kizirgulov, I. R., Mikolaychuk, O. V., Mikolaychuk, N. P. and Potapov, A. A., Phys. Rev. {\bf D 79}, 083006 (2009)

[8] Das Gupta, P., arXiv: 0905.1621v3 [gr-qc]

[9] Cowling, T. G., Astrophys. J. {\bf 114}, 272 (1951)

[10] Shu, F. H., Physics of Astrophysics, Vol.II, 1992 (University Science Books, California)

[11] Son, D. T., Phys. Rev. {\bf D 59}, 063008 (1999)

[12] Michel, F. C., Theory of Neutron Star Magnetospheres, 1991 (University of Chicago Press, Chicago)

[13] Kouveliotou, C. et. al, Nature {\bf 393}, 235 (1998)

[14] Goldreich, P. and Julian, W. H., Astrophys. J. {\bf 157}, 869 (1969)

[15] Mestel, L., Nature Phys. Sci. {\bf 233}, 149 (1971)

[16] Deutsch, A. J., Annales d' Astrophysique {\bf 18}, 1 (1955) 

[17] Kaburaki, O., Astrophys. Sp. Sci. {\bf 58}, 427 (1978)

[18] Cheng, K. S., Ho, C. and Ruderman, M.,, Astrophys. J. {\bf 300}, 500 (1986)

[19] Romani, R. W., Astrophys. J. {\bf 470}, 469 (1996)

[20] Takata, J., Shibata, S., Hirotani, K. and Chang, H. -K., Mon. Not. Roy. Astr. Soc. {\bf 366}, 1310 (2006)

[21] Alexander, S. H. S.  and Yunes, N., Phys. Rev. Lett. {\bf D 99}, 241101 (2007)

[22] Alexander, S. H. S.  and Yunes, N., Phys. Rev. {\bf D 75}, 124022 (2007)

[23] Grumiller, D.  and Yunes, N., Phys. Rev. {\bf D 77}, 044015 (2008)

\end{document}